\documentclass[%
 reprint,
 amsmath,amssymb,
 longbibliography,
 prb,
,superscriptaddress]{revtex4-2}

\usepackage[ruled]{algorithm2e}
\usepackage[noend]{algpseudocode}
\usepackage{type1cm} 
\usepackage{graphicx}
\usepackage{dcolumn}
\usepackage{bm}
\DeclareMathAlphabet{\mathpzc}{OT1}{pzc}{m}{it}
\usepackage{color}
\usepackage[usenames,dvipsnames,svgnames,table]{xcolor}
\usepackage{hyperref}

\usepackage{xcolor}

\hypersetup{pdfpagemode=FullScreen,colorlinks=true,linkcolor=NavyBlue,citecolor=BrickRed}

\begin{document}


\title{Spin-orbit coupling effects on orbital-selective correlations in a three-orbital model}

\author{Yin Chen}
\thanks{These authors contributed equally to this work.}
\affiliation{School of Physics and Beijing Key Laboratory of Opto-electronic Functional Materials $\&$ Micro-nano Devices, Renmin University of China, Beijing 100872, China}
\affiliation{Key Laboratory of Quantum State Construction and Manipulation (Ministry of Education), Renmin University of China, Beijing 100872, China}

\author{Yi-Heng Tian}
\thanks{These authors contributed equally to this work.}
\affiliation{School of Physics and Beijing Key Laboratory of Opto-electronic Functional Materials $\&$ Micro-nano Devices, Renmin University of China, Beijing 100872, China}
\affiliation{Key Laboratory of Quantum State Construction and Manipulation (Ministry of Education), Renmin University of China, Beijing 100872, China}

\author{Rong-Qiang He}\email{rqhe@ruc.edu.cn}
\affiliation{School of Physics and Beijing Key Laboratory of Opto-electronic Functional Materials $\&$ Micro-nano Devices, Renmin University of China, Beijing 100872, China}
\affiliation{Key Laboratory of Quantum State Construction and Manipulation (Ministry of Education), Renmin University of China, Beijing 100872, China}

\author{Zhong-Yi Lu}\email{zlu@ruc.edu.cn}
\affiliation{School of Physics and Beijing Key Laboratory of Opto-electronic Functional Materials $\&$ Micro-nano Devices, Renmin University of China, Beijing 100872, China}
\affiliation{Key Laboratory of Quantum State Construction and Manipulation (Ministry of Education), Renmin University of China, Beijing 100872, China}
\affiliation{Hefei National Laboratory, Hefei 230088, China}

\date{\today}

\begin{abstract}
In ruthenate materials, non-Fermi liquid (NFL) phases have been observed. We used the natural orbitals renormalization group (NORG) method as an impurity solver for dynamical mean-field theory (DMFT) to study a three-orbital Kanamori-Hubbard model with crystal field splitting, set at a specific filling of 2/3, which serves as a minimal Hamiltonian for the ruthenates. We find that without spin-orbit coupling (SOC), increasing the electron interactions results in an orbital-selective Mott (OSM) state, where the half-filled $d_{xy}$ orbital becomes a Mott insulator (MI) while the three-quarter-filled $d_{xz/yz}$ orbitals form a singular Fermi liquid (SFL). The OSM state is destroyed by the small SOC, which causes the small hybridization between the $d_{xy}$ and $d_{xz/yz}$ orbitals, resulting in both the orbitals exhibiting an NFL behavior. The $d_{xy}$ orbital is close to an MI and the $d_{xz/yz}$ orbitals are close to an SFL state. They exhibit distinct electronic scattering rates.
\end{abstract}


\maketitle



\section{Introduction}


Ruthenate systems exhibit a rich range of physical phenomena, such as unconventional superconductivity \cite{TISCOSR,ESCAMIR,MSCASECIR}, topological effects \cite{scaffidi2017weak,bienert2025s200a5m087,LRCDATOMIA}, unconventional metallic behavior \cite{LYJL,RTHHFLIC,NFLBICPB}, and orbital selectivity \cite{SCFHC,NOTMTIC,SCMF,CIC,BST,ORS,ECM,AOM,OOA,OSL,SOK}. For instance, Sr$_2$RuO$_4$ was widely studied as a candidate for spin-triplet superconductivity under ambient conditions, and its electronic properties can be altered by uniaxial pressure \cite{RIT}, exhibiting a non-Fermi liquid (NFL) behavior. The application of strain in Ba$_2$RuO$_4$ can cause an NFL state. An orbital-selective non-Fermi liquid (OSNFL) phase was observed in Ca$_{1.8}$Sr$_{0.2}$RuO$_4$ \cite{OSB,OSBO}.

In an octahedral crystal field, the five $d$ orbitals of a transition metal ion split into two distinct sets: the lower-energy $t_{2g}$ orbitals and the higher-energy $e_g$ orbitals. The $t_{2g}$ set consists of the $d_{xy}$, $d_{xz}$, and $d_{yz}$ orbitals, which are oriented between the coordinate axes. These orbitals are important for understanding the electronic behavior of ruthenate materials such as Sr$_2$RuO$_4$, Ba$_2$RuO$_4$, and Ca$_{1.8}$Sr$_{0.2}$RuO$_4$. In these ruthenate materials, each Ru atom is octahedrally coordinated by six oxygen atoms. In these materials, Ru has a valence of $4$. This coordination results in the $t_{2g}$ orbitals of Ru being partially occupied with $4$ electrons out of the maximum of $6$, giving a filling fraction of $2/3$. However, the distribution of these $4$ electrons among the $t_{2g}$ orbitals is not uniform in all cases. Specifically, in Ca$_{1.8}$Sr$_{0.2}$RuO$_4$, the $d_{xy}$ orbital is half-filled while the $d_{xz}$ and $d_{yz}$ orbitals are each occupied with $1.5$ electrons \cite{OSB}. The interplay among spin-orbit coupling (SOC), crystal-field splitting, and strong electron correlation in these materials gives rise to an NFL behavior.

To further theoretically investigate the NFL state in ruthenate systems, researchers have constructed a three-orbital Kanamori-Hubbard model with crystal field splitting and a filling of $2/3$ as a minimal model to study. A previous study \cite{OSBO} found that in this model, the orbital-selective Mott phase (OSMP) \cite{OSMITI}, where the $d_{xy}$ orbital is a Mott insulator and the $d_{xz/yz}$ orbitals are Fermi liquid (FL), can be altered by the SOC. This alteration results in the emergence of a new, small-region OSNFL phase. The study performed dynamical mean-field theory (DMFT) \cite{DMFTOS} calculations for the model by using exact diagonalization (ED) as the zero-temperature impurity solver. However, the ED method is limited by the number of bath sites that it can handle, which restricts the numerical accuracy of the DMFT approach. The existence of this small-region OSNFL phase in the model and whether or not the SOC can indeed alter the orbital-selective Mott (OSM) state remains uncertain and requires verification through higher-precision numerical methods.

In this paper, we employed the natural orbitals renormalization group (NORG) \cite{QRG,NOR,Wang2022solving} as the zero-temperature impurity solver for DMFT to study the three-orbital Kanamori-Hubbard model. In comparison to the ED method, the NORG approach can handle more bath sites, thus improving the numerical accuracy of DMFT. The obtained ground-state phase diagram of the model is presented in Fig.~\ref{fig:lmd-U-phase-graph}. Without SOC, we find that increasing the Hubbard interaction $U$ results in an orbital-selective Mott transition (OSMT). The half-filled $d_{xy}$ orbital becomes a Mott insulating state, while the three-quarter-filled $d_{xz/yz}$ orbitals form a singular Fermi liquid (SFL) \cite{RAS}. With small SOC, which causes the small hybridization between both the orbitals, the OSM state is altered \cite{ITOSMP,SOCES,TOFTSORPN}. The $d_{xy}$ and $d_{xz/yz}$ orbitals are in an NFL state, with the former being close to a Mott insulator and the latter close to an SFL state. Each set of orbitals has a different electronic scattering rate $\gamma \equiv -{\rm Im} \Sigma(\omega=0)$ \cite{NFLPALITSRIOTDHM}.


\section{Model and method}
This section details the theoretical model and numerical methodology employed to investigate the effects of SOC on orbital-selective correlations in a three-orbital system. The core of our study lies in a three-orbital Kanamori-Hubbard model that incorporates both local electronic interactions and SOC.

The Hamiltonian
\begin{equation}
\begin{aligned}
H&= t \sum_{\langle i j\rangle \alpha \sigma} C_{i \alpha \sigma}^{\dagger} C_{j \alpha \sigma}+\sum_{i \alpha \sigma}\left(\Delta_\alpha-\mu\right) n_{i \alpha \sigma} \\ & +U \sum_{i \alpha} n_{i \alpha \uparrow} n_{i \alpha \downarrow}+\left(U^{\prime}-J_z\right) \sum_{i \alpha>\beta \sigma} n_{i \alpha \sigma} n_{i \beta \sigma} \\ & +U^{\prime} \sum_{i \alpha>\beta \sigma} n_{i \alpha \sigma} n_{i \beta \bar{\sigma}}-J_f \sum_{i \alpha>\beta}\left[S_{i \alpha}^{+} S_{i \beta}^{-}+S_{i \alpha}^{-} S_{i \beta}^{+}\right] \\ & +J_p \sum_{i \alpha \neq \beta} C_{i \alpha \uparrow}^{\dagger} C_{i \alpha \downarrow}^{\dagger} C_{i \beta \downarrow} C_{i \beta \uparrow}+H_{\mathrm{SOC}}
\end{aligned} \label{eq:h1}
\end{equation}
of our model \cite{OSBO} consists of several terms, each corresponding to a specific type of contribution. These terms work together to describe the system's behavior. The first term is the kinetic energy term: $t \sum_{\langle i j\rangle \alpha \sigma} C_{i \alpha \sigma}^{\dagger} C_{j \alpha \sigma}$, which allows electrons to hop between the nearest-neighbor sites with an amplitude $t$. The operators $C_{i \alpha \sigma}^{\dagger}$ and $C_{i \alpha \sigma}$ respectively create and annihilate an electron at site $i$, with orbital index $\alpha$ and spin $\sigma$. Here, $\alpha$ corresponds to the three active $t_{2g}$ orbitals ($d_{xy}, d_{yz}, d_{xz}$). This term sets the stage for electron delocalization. Next, the term $\sum_{i \alpha \sigma}\left(\Delta_\alpha-\mu\right) n_{i \alpha \sigma}$ accounts for the energy level differences between the orbitals, where $\Delta_\alpha$ represents the onsite energy, $\mu$ is the chemical potential that controls the overall electron filling, and $n_{i \alpha \sigma}$ is the particle number operator, respectively.

Furthermore, the Hamiltonian includes local Coulomb interaction terms. The intra-orbital interaction term $U \sum_{i \alpha} n_{i \alpha \uparrow} n_{i \alpha \downarrow}$ describes the Coulomb repulsion between two electrons with opposite spins in the same orbital, with $U$ being the interaction strength. The inter-orbital interaction terms, given by $\left(U^{\prime}-J_z\right) \sum_{i \alpha>\beta \sigma} n_{i \alpha \sigma} n_{i \beta \sigma}+U^{\prime} \sum_{i \alpha>\beta \sigma} n_{i \alpha \sigma} n_{i \beta \bar{\sigma}}$, address the repulsion between electrons in different orbitals, with $U^{\prime}$ denoting the inter-orbital repulsion and $J_z$ being the Ising-type Hund's coupling. The spin-flip term $-J_f \sum_{i \alpha>\beta}\left[S_{i \alpha}^{+} S_{i \beta}^{-}+S_{i \alpha}^{-} S_{i \beta}^{+}\right]$ describes the process of flipping the spin of electrons in different orbitals with coupling strength $J_f$. The pair-hopping term, given by $J_p \sum_{i \alpha \neq \beta} C_{i \alpha \uparrow}^{\dagger} C_{i \alpha \downarrow}^{\dagger} C_{i \beta \downarrow} C_{i \beta \uparrow}$, describes the hopping of electron pairs between different orbitals, with coupling strength $J_p$. Lastly, the SOC term \cite{OSBO,MIT} 
\begin{equation}
\begin{aligned}
H_{\mathrm{SOC}}=\lambda \sum_{i \alpha \beta} \sum_{\sigma_1\sigma_2}\langle\alpha|\vec{L_i}|\beta\rangle\left\langle\sigma_1\right|\vec{S_i}\left|\sigma_2\right\rangle C_{i \alpha \sigma_1}^{\dagger} C_{i \beta \sigma_2}
\end{aligned} \label{eq:h2}
\end{equation}
introduces the coupling between electron spin and orbital angular momentum, with $\lambda$ representing the SOC strength. $\vec{L_i}$ and $\vec{S_i}$ are the orbital and spin angular momentum operators. 

To simplify the model and focus on essential physics, we introduce specific relationships between the parameters. To ensure the rotational invariance of the electronic interaction, we impose the condition $U = U^{\prime} + 2J_z$. We further assume an isotropic Hund's coupling, setting all components to be equal, i.e., $J_z = J_f = J_p = J = U/4$. This simplification reduces the number of independent parameters and allows us to focus on the overall effects of Hund’s coupling, without introducing additional complexities from different coupling strengths. 

To simulate crystal-field splitting, we set the orbital energies such that the $d_{yz}$ and $d_{xz}$ orbitals are degenerate, while the $d_{xy}$ orbital has a distinct energy level: $\Delta_{yz}=\Delta_{xz} \neq \Delta_{xy}$. This energy difference mimics the effect of a crystal field, which breaks the degeneracy of the $t_{2g}$-orbitals in a solid, leading to distinct energy levels. This splitting is essential for inducing an orbital-selective behavior in the system.

To represent a specific physical scenario, we set the orbital fillings so that the $d_{xy}$ orbital has $1$ electron, while the $d_{xz}$ and $d_{yz}$ orbitals each accommodate $1.5$ electrons. This filling configuration is chosen to match that of Ca$_{1.8}$Sr$_{0.2}$RuO$_{4}$ \cite{OSB} in the absence of SOC. The desired electron filling is achieved by adjusting the values of $\Delta_{\alpha}$ \cite{OSMTOOBDL}.

We employed the DMFT to solve this model on the Bethe lattice with infinite coordinations. The non-interacting density of states (DOSs) for the Bethe lattice is given by: $\rho_\alpha(\omega)=\frac{2}{\pi D^2} \sqrt{D^2-\omega^2}$, where $D$ is the half-bandwidth and serves as the energy unit. The core of the DMFT calculation is based on the self-consistency condition: $\Delta(\omega)=\frac{D^2}{4} G(\omega)$, where $\Delta(\omega)$ is the hybridization function and $G(\omega)$ is the local Green's function. In the previous studies \cite{OSBO}, the Weiss field was fitted with $6$ bath sites in the DMFT implementation. In our work, we fit the hybridization function with 24 bath sites, yielding accurate fitting and converged DMFT results with respect to the number of bath sites \cite{NFLAACWHDITBTOHM}. To solve the effective impurity problem within DMFT at zero temperature, we adopted the NORG method to solve the impurity model \cite{NFLAACWHDITBTOHM}. In our calculations of real-frequency Green's functions, we set the broadening factor to $\eta=0.02D$, which is important for obtaining accurate spectral functions. Additionally, a low-frequency cutoff was implemented by setting $\beta D=200$, where $\beta$ is the fictitious inverse temperature, effectively simulating the zero-temperature limit.


\section{Results}
\subsection{Ground-state phase diagram}

\begin{figure}[t!]
\includegraphics[width=8.6cm]{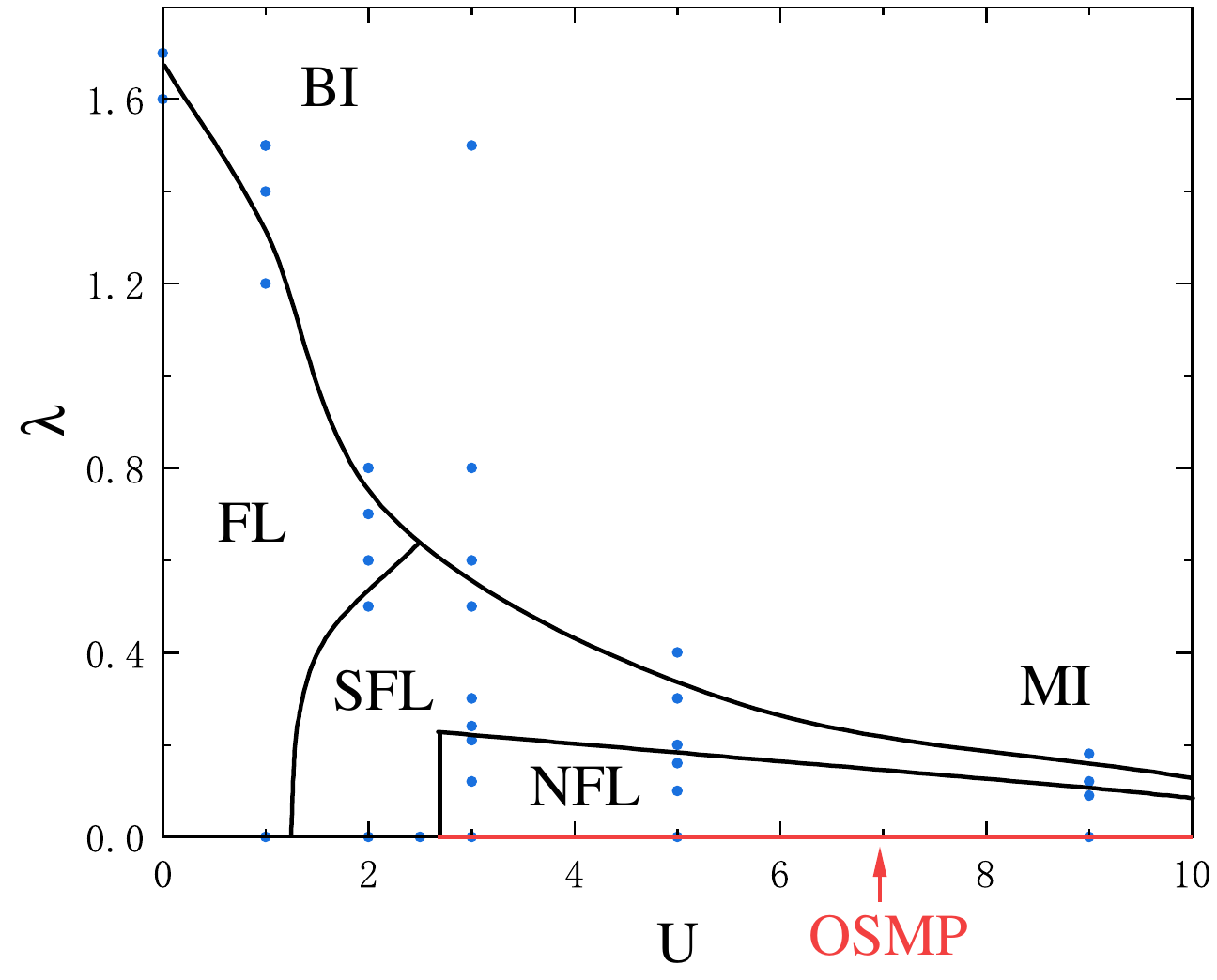}
\caption{Ground-state phase diagram for the three-orbital Kanamori-Hubbard model in the plane of $\lambda$ and $U$. It includes phases FL, SFL, OSMP, NFL, MI, and BI, where MI denotes Mott insulator and BI denotes band insulator. The blue dots represent the parameter points for which the phases have been explicitly determined through DMFT calculations.}\label{fig:lmd-U-phase-graph}
\end{figure}

Based on how the Matsubara self-energies change with Coulomb interaction $U$ and SOC strength $\lambda$, we construct a ground-state phase diagram (Fig.~\ref{fig:lmd-U-phase-graph}), which features phases including FL, SFL, OSMP, NFL, MI, and BI. Here, MI stands for Mott insulator and BI for band insulator. The OSMP is destroyed by the introduction of SOC, which causes the hybridization between the $d_{xy}$ and $d_{xz/yz}$ orbitals, resulting in that both the orbitals exhibit metallic properties. Both FL and SFL have a vanishing electronic scattering rate. For FL, ${\rm Im}\Sigma(i\omega_n)$ exhibits linearity near zero frequency, whereas for SFL, ${\rm Im}\Sigma(i\omega_n)$ shows nonlinearity near zero frequency. For NFL, the electronic scattering rate is nonzero.

\subsection{OSMT in the absence of SOC}

\begin{figure}[t!]
\includegraphics[width=8.6cm]{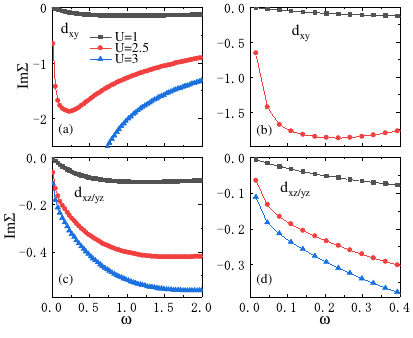}
\caption{${\rm Im}\Sigma(i\omega_n)$ of the $d_{xy}$ and $d_{xz/yz}$ orbitals at $U = 1,2.5,3$ and $\lambda = 0$. As $U$ increases, the ground state transitions through different phases. When $U = 1$, it is in an FL phase. When $U = 2.5$, it is in an SFL phase. When $U = 3$, it is in an OSMP.}\label{fig:seimp-U1-U2d5-U3}
\end{figure}

\begin{figure}[t!]
\includegraphics[width=8.6cm]{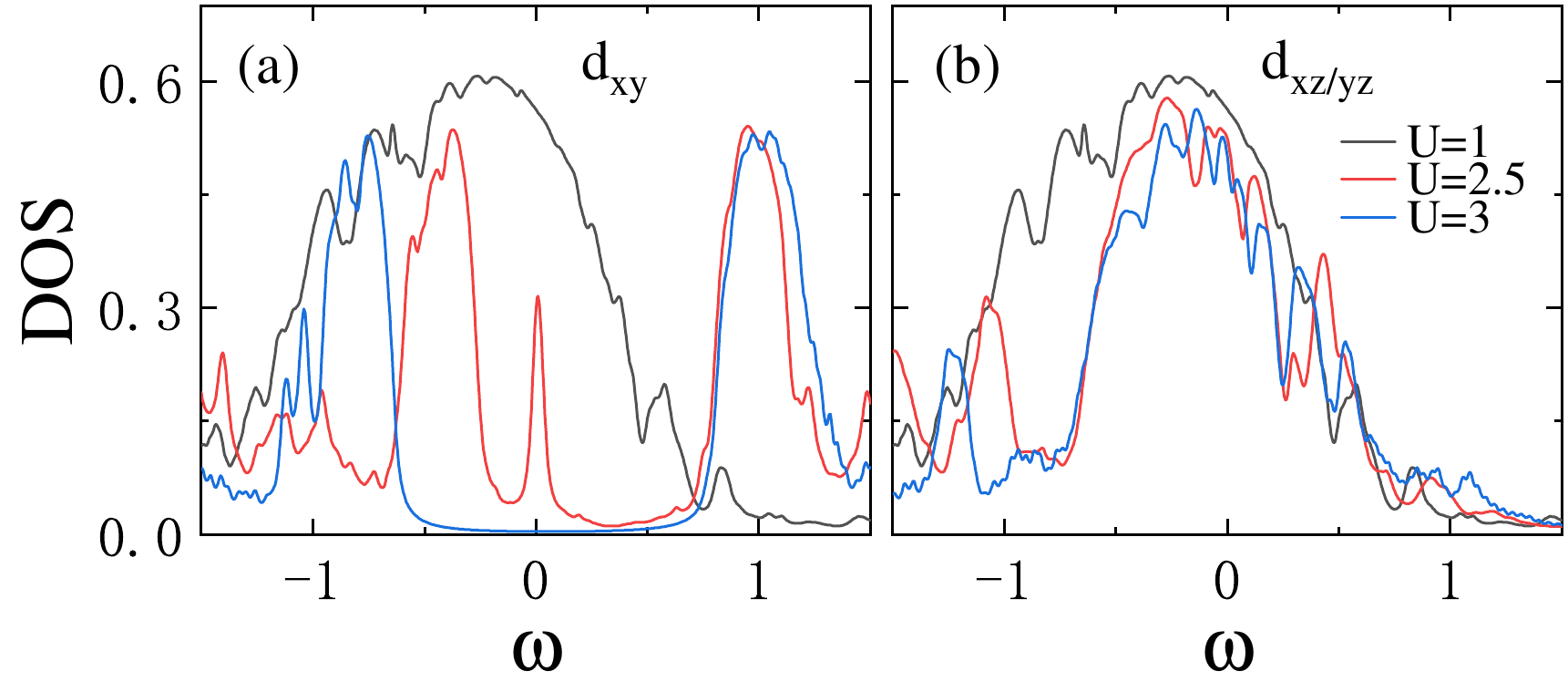}
\caption{DOS of the $d_{xy}$ and $d_{xz/yz}$ orbitals at $U = 1,2.5,3$ and $\lambda = 0$.}\label{fig:dos-U1-U2d5-U3}
\end{figure}

\begin{figure}[t!]
\includegraphics[width=8.6cm]{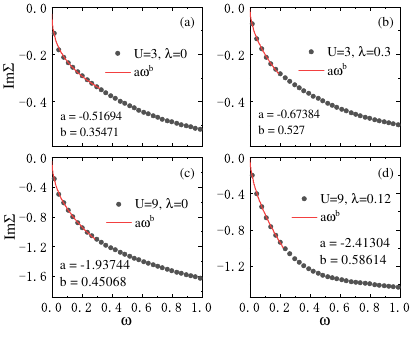}
\caption{${\rm Im}\Sigma(i\omega_n)$ of the $d_{xz/yz}$ orbitals at $U=3$ with $\lambda=0$ and $0.3$, and at $U=9$ with $\lambda=0$ and $0.12$. We fit the low-frequency region of ${\rm Im}\Sigma(i\omega_n)$ with a power-law function $a \omega^b$. ${\rm Im}\Sigma(i\omega_n)$ of the $d_{xz/yz}$ orbitals shows nonlinearity at low frequencies, indicating that the $d_{xz/yz}$ orbitals are in an SFL state.}\label{fig:seimp-U3-lmd0-0d3-U9-lmd0-0d12}
\end{figure}

In the absence of SOC, as the Coulomb interaction strength $U$ increases, the ground state transitions through different phases. Initially, at a lower $U$, such as $U = 1$, the ground state is in an FL phase. For the $d_{xy}$ and $d_{xz/yz}$ orbitals, ${\rm Im}\Sigma(i\omega_n)$ all exhibit linearity near zero frequency (Fig.~\ref{fig:seimp-U1-U2d5-U3}), and they have a vanishing electronic scattering rate.

As $U$ increases further, both the orbitals transition from an FL state to an SFL state. For example, when $U = 2.5$, the ground state is in the SFL phase. Both the orbitals have a vanishing electronic scattering rate, and their ${\rm Im}\Sigma(i\omega_n)$ exhibit nonlinearity near zero frequency (Fig.~\ref{fig:seimp-U1-U2d5-U3}).

Finally, at sufficiently large $U$, the ground state undergoes an OSMT, with the $d_{xy}$ orbital becoming Mott insulating while the $d_{xz/yz}$ orbitals still being of SFL. For example, when $U = 3$, the ground state is in the OSMP. The half-filled $d_{xy}$ orbital becomes an MI, exhibiting a downward divergence of ${\rm Im}\Sigma(i\omega_n)$ as the frequency approaches zero (Fig.~\ref{fig:seimp-U1-U2d5-U3}(a)). Correspondingly, the DOS shows the lower and upper Hubbard bands on both sides of the Fermi level, with a gap separating them (Fig.~\ref{fig:dos-U1-U2d5-U3}(a)). We fit the low-frequency region of ${\rm Im}\Sigma(i\omega_n)$ of the $d_{xz/yz}$ orbitals to a power-law function $a \omega^b$. The fitting parameters for the $d_{xz/yz}$ orbitals are $b=0.35471$ for $U=3$ (Fig.~\ref{fig:seimp-U3-lmd0-0d3-U9-lmd0-0d12}(a)) and $b=0.45068$ for $U=9$ (Fig.~\ref{fig:seimp-U3-lmd0-0d3-U9-lmd0-0d12}(c)). ${\rm Im}\Sigma(i\omega_n)$ of the $d_{xz/yz}$ orbitals has a vanishing electronic scattering rate and exhibits nonlinearity at low frequencies, indicating that the $d_{xz/yz}$ orbitals stay in an SFL state (Figs.~\ref{fig:seimp-U1-U2d5-U3}(c, d)).

\subsection{SOC effects on orbital-selective Mott physics}

\begin{figure}[t!]
\includegraphics[width=8.6cm]{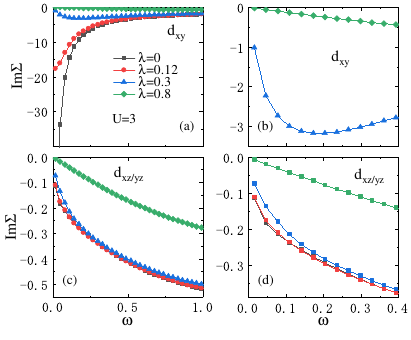}
\caption{${\rm Im}\Sigma(i\omega_n)$ of the $d_{xy}$ and $d_{xz/yz}$ orbitals at $\lambda = 0, 0.12, 0.3, 0.8$ and $U = 3$. As $\lambda$ increases, the ground state transitions through different phases. When $\lambda = 0$, it is in an OSMP. When $\lambda = 0.12$, it is in an NFL phase. When $\lambda = 0.3$, it is in an SFL phase. When $\lambda = 0.8$, it is in a BI phase.}\label{fig:seimp-U3-lmd0-0d12-0d3-0d8}
\end{figure}

\begin{figure}[t!]
\includegraphics[width=8.6cm]{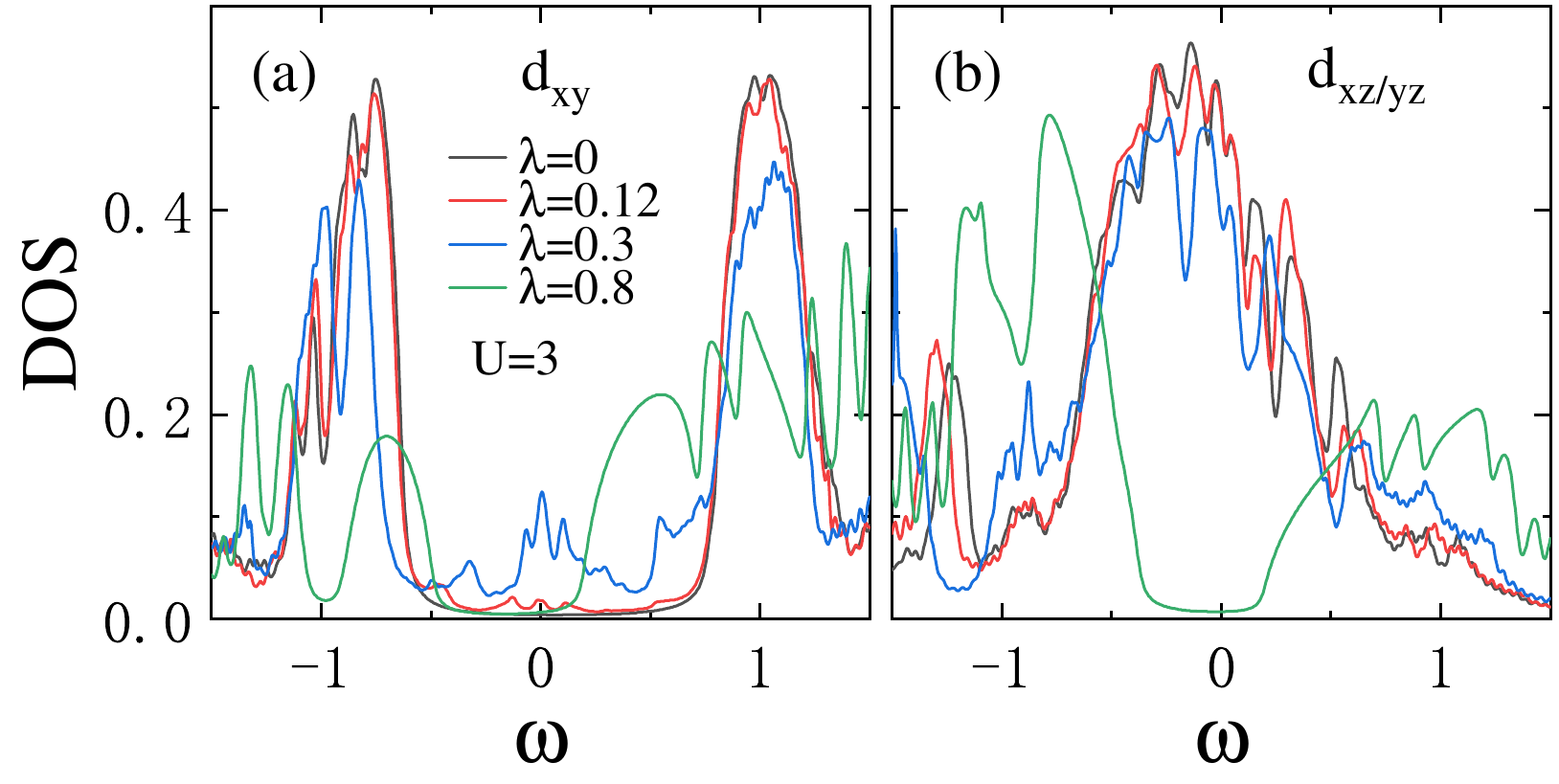}
\caption{DOS of the $d_{xy}$ and $d_{xz/yz}$ orbitals at $\lambda = 0,0.12,0.3,0.8$ and $U = 3$.}\label{fig:dos-U3-lmd0-0d12-0d3-0d8}
\end{figure}

\begin{figure}[t!]
\includegraphics[width=8.6cm]{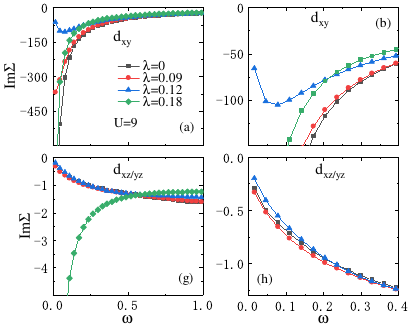}
\caption{${\rm Im}\Sigma(i\omega_n)$ of the $d_{xy}$ and $d_{xz/yz}$ orbitals at $\lambda = 0, 0.09, 0.12, 0.18$ and $U = 9$. As $\lambda$ increases, the ground state transitions through different phases. When $\lambda = 0$, it is in an OSMP. When $\lambda = 0.09$, it is in an NFL phase. When $\lambda = 0.12$, it is in an SFL phase. When $\lambda = 0.18$, it is in an MI phase.}\label{fig:seimp-U9-lmd0-0d09-0d12-0d18}
\end{figure}

\begin{figure}[t!]
\includegraphics[width=8.6cm]{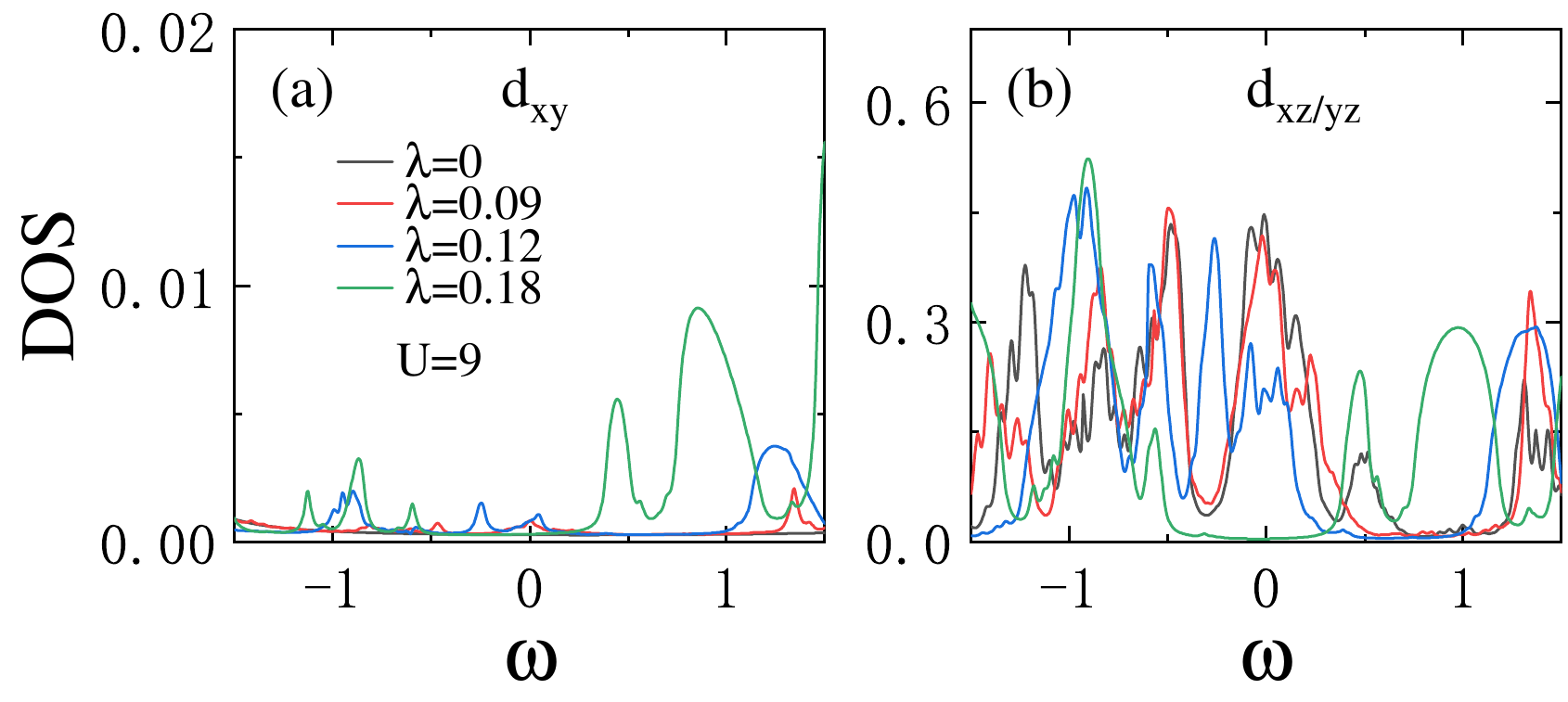}
\caption{DOS of the $d_{xy}$ and $d_{xz/yz}$ orbitals at $\lambda = 0,0.09,0.12,0.18$ and $U = 9$.}\label{fig:dos-U9-lmd0-0d09-0d12-0d18}
\end{figure}

In the absence of SOC, we find that increasing $U$ results in an OSMT. In this OSMP, the $d_{xy}$ orbital becomes an MI and the $d_{xz/yz}$ orbitals form an SFL. For example, when $U = 3,\lambda = 0$, or when $U = 9,\lambda = 0$, the ground state is in an OSMP.

As $\lambda$ increases, the ground state transitions through different phases. Initially, at a lower $\lambda$, such as $U = 3,\lambda = 0.12$ or $U = 9,\lambda = 0.09$, the ground state is in an NFL phase. The $d_{xy}$ orbital, initially in a Mott state without SOC, becomes a weakly metallic state. This transition is due to the small hybridization between the $d_{xy}$ and $d_{xz/yz}$ orbitals \cite{ITOSMP,SOCES,TOFTSORPN}, which is induced by the SOC. The $d_{xy}$ orbital exhibits a large electronic scattering rate (Figs.~\ref{fig:seimp-U3-lmd0-0d12-0d3-0d8}(a) and ~\ref{fig:seimp-U9-lmd0-0d09-0d12-0d18}(a)), accompanied by a small DOS peak at the Fermi level (Figs.~\ref{fig:dos-U3-lmd0-0d12-0d3-0d8}(a) and ~\ref{fig:dos-U9-lmd0-0d09-0d12-0d18}(a)). The $d_{xy}$ orbital is in an NFL state. Meanwhile, the $d_{xz/yz}$ orbitals, initially in an SFL state, transition into an NFL state due to the hybridization. The electronic scattering rate of the $d_{xz/yz}$ orbitals changes from initially vanishing to a small value (Figs.~\ref{fig:seimp-U3-lmd0-0d12-0d3-0d8} and ~\ref{fig:seimp-U9-lmd0-0d09-0d12-0d18}). For more details on the small electronic scattering rate of the $d_{xz/yz}$ orbitals, please refer to the Appendix.

The electronic scattering rate of the $d_{xz/yz}$ orbitals is very small and may be difficult to observe experimentally. The $d_{xz/yz}$ orbitals are close to an SFL state. In contrast, the scattering rate of the $d_{xy}$ orbital is large. The $d_{xy}$ orbital is close to an MI.

With further increase of $\lambda$, the hybridization is enhanced, and both the orbitals transition into an SFL state. For example, when $U = 3,\lambda = 0.3$, or when $U = 9,\lambda = 0.12$, the ground state is in the SFL phase. We fit the low-frequency region of ${\rm Im}\Sigma(i\omega_n)$ of the $d_{xz/yz}$ orbitals to a power-law function $a \omega^b$. The fitting parameters for the $d_{xz/yz}$ orbitals are $b=0.527$ for $U=3, \lambda = 0.3$ (Fig.~\ref{fig:seimp-U3-lmd0-0d3-U9-lmd0-0d12}(b)) and $b=0.58614$ for $U=9, \lambda = 0.12$ (Fig.~\ref{fig:seimp-U3-lmd0-0d3-U9-lmd0-0d12}(d)). ${\rm Im}\Sigma(i\omega_n)$ of the $d_{xz/yz}$ orbitals has a vanishing electronic scattering rate at zero frequency and exhibits nonlinearity at low frequencies, indicating that the $d_{xz/yz}$ orbitals stay in an SFL state (Figs.~\ref{fig:seimp-U3-lmd0-0d12-0d3-0d8} and ~\ref{fig:seimp-U9-lmd0-0d09-0d12-0d18}).

\subsection{Insulating phases driven by SOC}

At large SOC strength $\lambda$, both the orbitals transition into an insulating state. The nature of the insulating phases depends on $U$; the ground state is in a BI phase for small $U$, and in an MI phase for large $U$.

In the BI phase, the SOC dominates and induces a band insulating behavior. Specifically, the occurrence of the BI phase at small $U$ can be understood within the $|J, \pm m\rangle$ basis, where the local Hamiltonian Eq.(\ref{eq:h2}) can be diagonalized. The SOC raises the energies of the $\left|\frac{1}{2}, \pm \frac{1}{2}\right\rangle$ bands to approximately $\lambda$, and lowers those of the $\left|\frac{3}{2}, \pm \frac{1}{2}\right\rangle$ bands and $\left|\frac{3}{2}, \pm \frac{3}{2}\right\rangle$ bands to about $-\frac{\lambda}{2}$ \cite{MIT,OSBO}. With a total of four electrons, the $\left|\frac{3}{2}, \pm \frac{3}{2}\right\rangle$ and $\left|\frac{3}{2}, \pm \frac{1}{2}\right\rangle$ bands become fully filled, while the $\left|\frac{1}{2}, \pm \frac{1}{2}\right\rangle$ bands remain empty. This band configuration underlies the emergence of the BI phase.

\begin{figure}[t!]
\includegraphics[width=8.6cm]{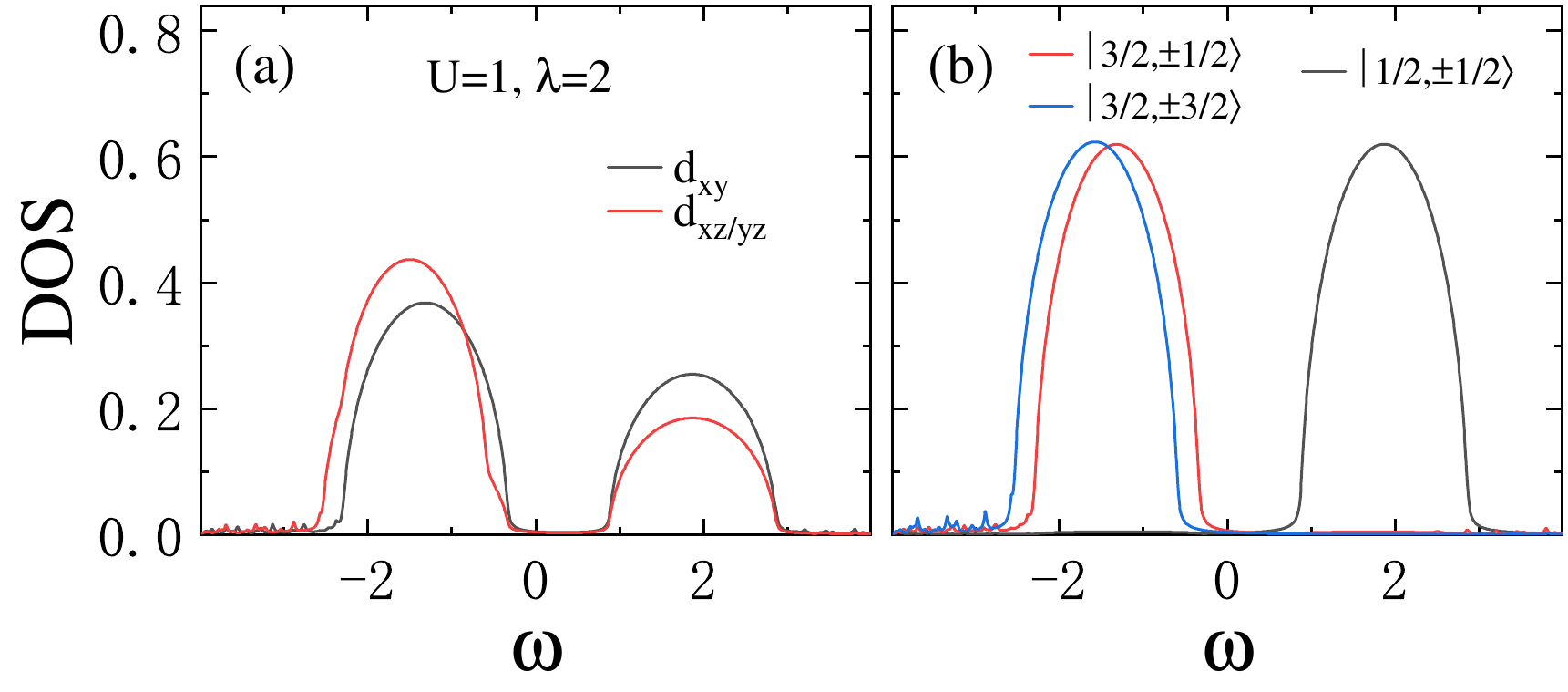}
\caption{(a) DOS of the $d_{xy}$ and $d_{xz/yz}$ orbitals at $U = 1,\lambda = 2$. (b) DOS of the $\left|\frac{1}{2}, \pm \frac{1}{2}\right\rangle$, $\left|\frac{3}{2}, \pm \frac{1}{2}\right\rangle$, and $\left|\frac{3}{2}, \pm \frac{3}{2}\right\rangle$ orbitals at $U = 1,\lambda = 2$.}\label{fig:dos-U1-lmd2}
\end{figure}

To be specific, when $U = 1,\lambda = 2$, the ground state is in the BI phase. The occupation numbers are as follows: the $\left|\frac{1}{2}, \pm \frac{1}{2}\right\rangle$ bands have an occupation of $0.058$, the $\left|\frac{3}{2}, \pm \frac{1}{2}\right\rangle$ bands have an occupation of $1.946$, and the $\left|\frac{3}{2}, \pm \frac{3}{2}\right\rangle$ bands have an occupation of $1.994$. The DOS of the $\left|\frac{1}{2}, \pm \frac{1}{2}\right\rangle$ bands shows a peak lies above the Fermi level, while the $\left|\frac{3}{2}, \pm \frac{1}{2}\right\rangle$ and $\left|\frac{3}{2}, \pm \frac{3}{2}\right\rangle$ bands are below the Fermi level, which is consistent with their high occupation (Fig.~\ref{fig:dos-U1-lmd2}(b)).  The DOS for both the orbitals is zero at zero frequency (Fig.~\ref{fig:dos-U1-lmd2}(a)).

The previous study \cite{OSBO} has explained the SOC-assisted Mott phase. SOC enhances band polarization and leads to the full filling of the $\left|\frac{3}{2}, \pm \frac{3}{2}\right\rangle$ bands. The remaining two electrons then reside in the $\left|\frac{3}{2}, \pm \frac{1}{2}\right\rangle$ and $\left|\frac{1}{2}, \pm \frac{1}{2}\right\rangle$ bands, resulting in an effective half-filled system, instead of the original four electrons distributed across these bands. For example, when $U = 9,\lambda = 0.18$, the ground state is in an MI phase. The occupation numbers are as follows: the $\left|\frac{1}{2}, \pm \frac{1}{2}\right\rangle$ orbitals have an occupation of $0.896$, the $\left|\frac{3}{2}, \pm \frac{1}{2}\right\rangle$ orbitals have an occupation of $1.108$, and the $\left|\frac{3}{2}, \pm \frac{3}{2}\right\rangle$ orbitals have an occupation of $1.992$. For both the orbitals, the DOS are zero at the Fermi level (Fig.~\ref{fig:dos-U9-lmd0-0d09-0d12-0d18}), and ${\rm Im}\Sigma(i\omega_n)$ diverges downward as the frequency approaches zero (Fig.~\ref{fig:seimp-U9-lmd0-0d09-0d12-0d18}).

There is no clear boundary between the BI phase and the MI phase; it is merely a crossover. Across this crossover (as $U$ increases), the charge excitation gap does not close, and the band occupancies change continuously. For example, at $U=3,\lambda=0.8$, the DOS for both the orbitals are zero at the Fermi level (Fig.~\ref{fig:dos-U3-lmd0-0d12-0d3-0d8}). The occupancy of the $\left|\frac{1}{2}, \pm \frac{1}{2}\right\rangle$ bands is $0.388$, which is well between completely empty and half-filled, while the occupancy of the $\left|\frac{3}{2}, \pm \frac{3}{2}\right\rangle$ bands is $1.647$, falling between half-filled and fully filling.

\section{Summary}
We employed NORG method as the zero-temperature impurity solver for DMFT to study the three-orbital Kanamori-Hubbard model. Compared to the ED method, the NORG approach can handle more bath sites, thereby enhancing the numerical accuracy of DMFT. The ground-state phase diagram of the model was then obtained (Fig.~\ref{fig:lmd-U-phase-graph}).

Without SOC, increasing $U$ leads to an OSMT, in which the half-filled $d_{xy}$ orbital becomes an MI, while the three-quarter-filled $d_{xz/yz}$ orbitals enter an SFL state. In this SFL state, ${\rm Im}\Sigma(i\omega_n)$ exhibits nonlinearity near zero frequency, and the electronic scattering rate vanishes.

Introducing the small SOC alters the original OSM state. Both the orbitals enter an NFL state. The $d_{xy}$ orbital is close to a Mott insulator, and the $d_{xz/yz}$ orbitals are close to an SFL state. Each set of orbitals has a different electronic scattering rate. With further increase of $\lambda$, both the orbitals transition from the NFL state into an SFL state. Finally, with increasing $\lambda$, both the orbitals enter insulating states. Their nature is determined by $U$: a BI phase emerges at small $U$, whereas an MI phase appears at large $U$. There is no clear phase boundary separating the BI phase and the MI phase.

\section*{acknowledgments}
This work was supported by the National Key R\&D Program of China (Grants No. 2024YFA1408601 and No. 2024YFA1408602), the National Natural Science Foundation of China (Grant No. 12434009), and the Innovation Program for Quantum Science and Technology (Grant No. 2021ZD0302402). Computational resources were provided by the Physical Laboratory of High Performance Computing in Renmin University of China.

\section*{Appendix: small electronic scattering rate of the $d_{xz/yz}$ orbitals in the NFL phase}
\addcontentsline{toc}{section}{Appendix: small electronic scattering rate of the $d_{xz/yz}$ orbitals in the NFL phase}
\label{appB}

In Fig.~\ref{fig:seimp-U3-lmd0-0d12-0d3-0d8}(d), it can be seen that for $U=3, \lambda=0.12$, the value of ${\rm Im}\Sigma(i\omega_n)$ at the lowest frequency (closest to zero frequency) is almost identical to that for $U=3, \lambda=0$. However, at the second-lowest frequency point, the value of ${\rm Im}\Sigma(i\omega_n)$ for $U=3, \lambda=0.12$ is notably higher than that for $U=3, \lambda=0$, suggesting a small intercept at zero frequency for the $d_{xz/yz}$ orbitals at $U=3, \lambda=0.12$.

In Fig.~\ref{fig:seimp-U9-lmd0-0d09-0d12-0d18}(d), for $U=9, \lambda=0.09$, the value of ${\rm Im}\Sigma(i\omega_n)$ at the second-lowest frequency point is very close to that for $U=9, \lambda=0$. However, at the lowest frequency point, the value for $U=9, \lambda=0.09$ lies notably below that for $U=9, \lambda=0$, indicating a small intercept at zero frequency for the $d_{xz/yz}$ orbitals at $U=9, \lambda=0.09$.

In the NFL phase, the $d_{xz/yz}$ orbitals exhibit a small electronic scattering rate.

\bibliography{soc3orb}

\end{document}